\documentclass[twocolumn,superscriptaddress,floatfix,aps,prl]{revtex4}
\usepackage{graphicx,subfigure,epsfig}
\usepackage{dcolumn}
\usepackage{amssymb}
\usepackage{times}
\usepackage{amsmath}
\usepackage{amsfonts}
\usepackage{mathrsfs}
\usepackage{setspace}
\usepackage{latexsym}
\usepackage{bbm}
\usepackage{float}
\usepackage{flafter}
\usepackage{bm}
\usepackage{epstopdf}
\usepackage{hyperref}
\usepackage{color}
\usepackage{multirow}

\begin{document}

\title{Robust non-Abelian spin liquid and possible intermediate phase in antiferromagnetic Kitaev model with magnetic field}

\author{Zheng Zhu}
\affiliation{Department of Physics, Massachusetts Institute of Technology, Cambridge, MA, 02139, USA}
\author{Itamar Kimchi}
\affiliation{Department of Physics, Massachusetts Institute of Technology, Cambridge, MA, 02139, USA}
\author{D. N. Sheng}
\affiliation{Department of Physics and Astronomy, California State University, Northridge, CA, 91330, USA}
\author{Liang Fu}
\affiliation{Department of Physics, Massachusetts Institute of Technology, Cambridge, MA, 02139, USA}

\begin{abstract}
We investigate the non-Abelian topological chiral spin liquid phase in the two-dimensional (2D) Kitaev honeycomb model subject to a magnetic field. By combining density matrix renormalization group (DMRG) and exact diagonalization (ED) we study the energy spectra, entanglement, topological degeneracy, and expectation values of Wilson loop operators, allowing for robust characterization.
While the ferromagnetic (FM) Kitaev spin liquid is already destroyed by a weak magnetic field with Zeeman energy $H_*^\text{FM} \approx 0.02$, the antiferromagnetic (AFM) spin liquid remains robust up to a magnetic field that is an order of magnitude larger, $H_*^\text{AFM} \approx 0.2$. Interestingly, for larger fields $H_*^\text{AFM} < H < H_{**}^\text{AFM}$, an intermediate gapless phase is observed, before a second transition to the high-field partially-polarized paramagnet. We attribute this rich phase diagram, and the remarkable stability of the chiral topological phase in the AFM Kitaev model, to the interplay of strong spin-orbit coupling and frustration enhanced by the magnetic field. Our findings suggest relevance to recent experiments on RuCl$_3$ under magnetic fields.
\end{abstract}

\maketitle

\emph{Introduction.} The search for highly entangled quantum states of matter such as quantum spin liquids (QSLs)  has intensified in recent years ~\cite{Wen2004,Lee2008,Balents,Zhou2017}. The peculiarity of QSLs lies not only in the  absence of  magnetic long-rang order even at zero temperature, but more importantly in exhibiting fractionalized excitations and topological ground state degeneracy. Among various theoretically proposed QSLs,  a remarkable example is the Kitaev model of spins with nearest-neighbor interactions on the two-dimensional (2D) honeycomb lattice \cite{Kitaev2006}. This model is solved exactly by mapping it into a model of Majorana fermions coupled to an emergent static $\mathbb{Z}_2$ gauge field. The ground state is  either a gapless spin liquid  or, with weak time reversal breaking, a gapped spin liquid phase. The latter harbors a non-Abelian anyon known as an Ising anyon, a descendant of vortices in two-dimensional $p+ip$ superconductors \cite{Read2000}. The exact solution of the apparently simple Kitaev model has motivated a search for the physical realization of non-Abelian QSL \cite{Rau2016,Trebst2017,Hermanns2017}.

\begin{figure}[tp]
\begin{center}
\includegraphics[width=0.38\textwidth]{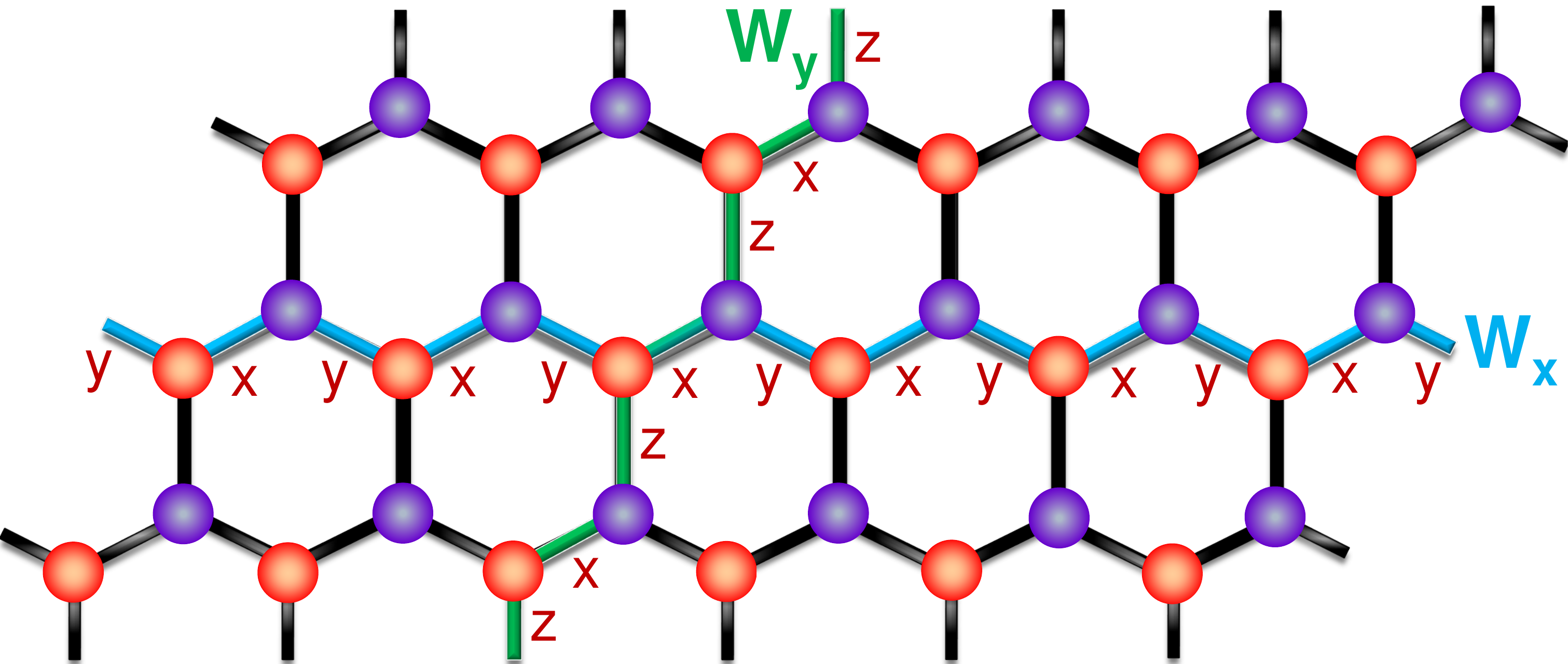}
\end{center}
\par
\renewcommand{\figurename}{Fig.}
\caption{(Color online) The honeycomb lattice and Wilson loop operators. The honeycomb lattice is spanned by unit vectors $(1,0)$ and $(1/2, \sqrt{3}/2)$ with lattice size $N=L_x\times L_y \times 2$. Green and blue loops denote Wilson loop operators along vertical and horizontal periodic boundary conditions on the torus, respectively.}
\label{Fig1}
\end{figure}

\begin{figure*}[btp]
\begin{center}
\includegraphics[width=1.0\textwidth]{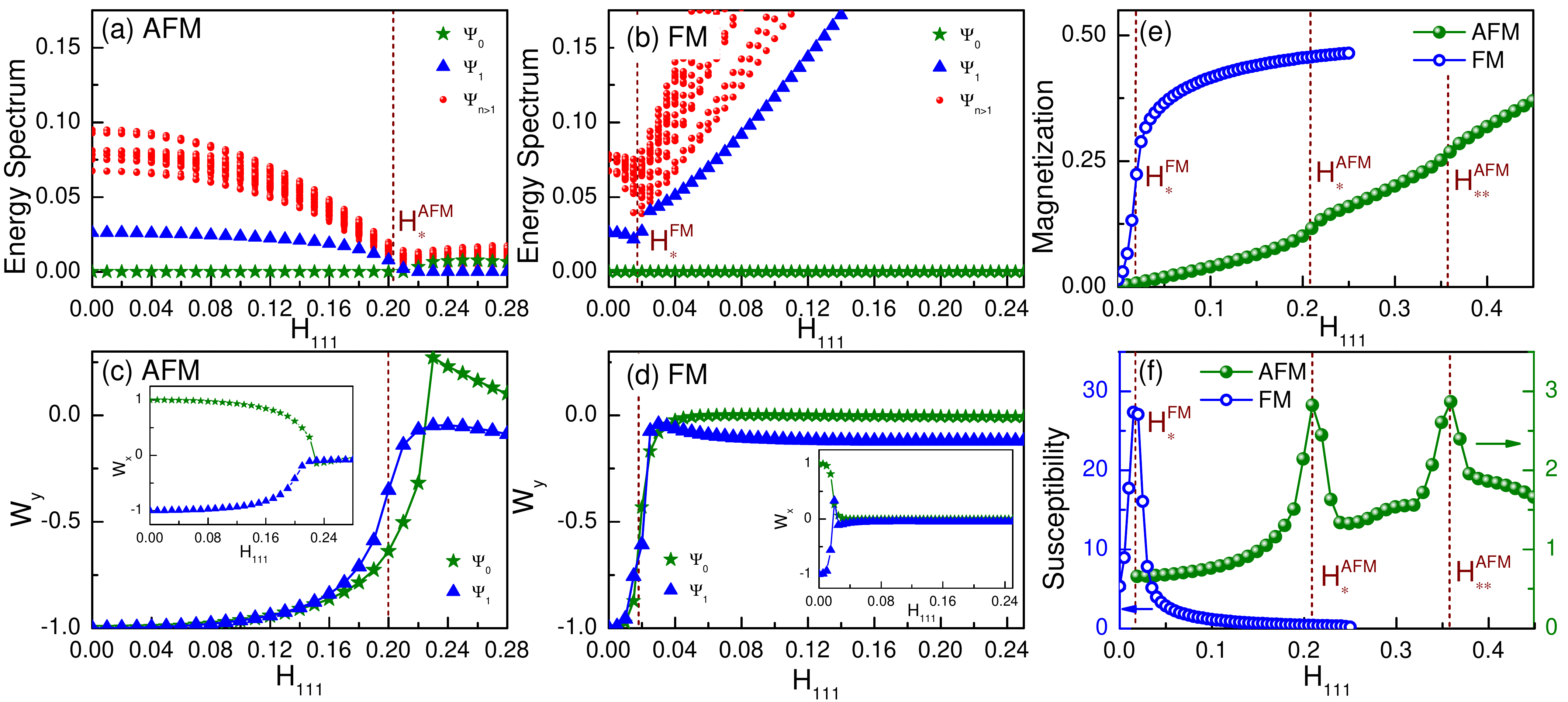}
\end{center}
\par
\renewcommand{\figurename}{Fig.}
\caption{(Color online) Panels (a, b): For the AFM (a) and FM (b) Kitaev models in a magnetic field, the pair of topological ground states (approximately degenerate on this $N= 4\times 3 \times2$ torus) are separated from higher energy states by an energy gap, within the topological phase $H<H_{*}$ where $H_*^\text{AFM} \approx 0.2$ and $H_*^\text{FM} \approx 0.02$. Panels (c, d): Wilson loop operators $W_y$ (main panel) and $W_x$ (inset) for the two lowest energy states, for AFM (c) and FM (d) models. The two states have $W_y=-1$ but are distinguished by $W_x=\pm1$.
Panel (e): The magnetization curves show the transitions and the stark difference between the AFM and FM models.
Panel (f):  The second order derivative of ground state energy with respect to field, or equivalently the magnetic susceptibility: note the difference in magnitudes between AFM and FM models. In the AFM case, the transition to the polarized high-field phase is achieved only at a second peak with $H_{**} \approx 0.36$.}
\label{Fig2_ED}
\end{figure*}

The highly anisotropic and spatially dependent spin interaction in the Kitaev model can conceivably appear in Mott insulators with strong spin-orbit coupling and $j=1/2$ local moments. In particular Jackeli and  Khaliullin \cite{Jackeli2009} proposed a mechanism for Kitaev interaction in transition metal oxides with edge-sharing oxygen octahedra.
By now, in addition to various  three-dimensional compounds\cite{Iridates4,Iridates6,Iridates1}, a variety of two dimensional layered honeycomb lattice magnets\cite{Chaloupka2010} have been discovered, including  Na$_2$IrO$_3$ \cite{Singh2010,Singh2012,Choi2012},  $\alpha$-Li$_2$IrO$_3$ \cite{Singh2010,Singh2012}, a hydrated variant H$_3$LiIr$_2$O$_6$\cite{Takagi2017}, and RuCl$_3$ \cite{Plumb2014}.

Aside from the spin liquid candidates Na$_4$Ir$_3$O$_8$ (hyperkagome\cite{Iridates1,Kimchi2014}) and H$_3$LiIr$_2$O$_6$ (honeycomb\cite{Takagi2017}),  these compounds are magnetically ordered at sufficiently low temperatures, indicating the presence of additional spin interactions beyond the Kitaev model.
Nonetheless, various experimental and theoretical works suggest the magnetic ordered states are proximate to a spin-liquid phase\cite{Hermanns2017,Gohlke2017,Gohlke2017a,Katukuri2014,Banerjee2016,Winter2017,Janssen2017,Banerjee2017,Banerjee2017b,Ponomaryov2017,Little2017,ZWang2017, Song2016}.
To understand the nature of quantum phases realized in materials, it is helpful to compare experimental findings with expected signatures of perturbed Kitaev Hamiltonians. However,even with the large body of available experimental data, the sign of the Kitaev interaction in the honeycomb magnets remains an open question  \cite{Ran2017,Wang2017,Banerjee2016,Winter2017,Janssen2017,Banerjee2017,Banerjee2017b,Ponomaryov2017,Little2017,Kim2015,Rau2014,Yamaji2014,Chaloupka2015,Rousochatzakis2015,Sizyuk2016,Kim2016,Chaloupka2016,Winter2016}.
In models with strong spin orbit coupling, the Curie-Weiss temperature may not reflect the dominant interaction due to cancellation among the various spin-orbit-coupled exchanges; for instance $T_{CW}$ may even vanish\cite{Reuther2011}.
Interestingly for RuCl$_3$ it has recently been argued that natural models with nearest-neighbor exchanges involve strong $\Gamma$ exchange or have dominant likely \textit{antiferromagnetic} Kitaev exchange \cite{Chaloupka2013,Janssen2017,Banerjee2016}.

For the pure Kitaev model, the different signs of the Kitaev exchange are related by a sublattice dependent  transformation, leading to identical energy spectrum. However, under an external magnetic field, ferromagnetic (FM) and antiferromagnetic (AFM) Kitaev models are expected to behave differently. Previous theoretical studies mainly focused on the FM Kitaev model, and found that the non-Abelian spin liquid phase only survives up to a very small magnetic field $H_*^\text{FM}\thickapprox 0.02$ by Jiang \emph{et al.}\cite{Jiang2011}. To our knowledge,  except semiclassically \cite{Janssen2016}, the AFM Kitaev model in a magnetic field has not yet been studied.

In this Letter, we study the AFM Kitaev model in a magnetic field by performing extensive exact diagonalization (ED) and density matrix renormalization group (DMRG) simulations. The energy spectra, the expectation value of Wilson loop operator and the ground state degeneracy as a function of the magnetic field are computed and compared with exact analytical results at zero field. We find the presence of the non-Abelian QSL phase in the AFM Kitaev model over a wide range of magnetic field up to $H_*^\text{AFM} \thickapprox 0.2$, an order-of-magnitude larger than that of the FM Kitaev model.  Moreover, before a second transition at $H_{**}^\text{AFM}\approx0.36$ to the high-field partially-polarized paramagnet , an intermediate gapless phase is observed  for fields $H_*^\text{AFM} < H < H_{**}^\text{AFM}$.

 \emph{Model and Method.}---We consider the Kitaev honeycomb model subject to an external magnetic field $\mathbf{H}$  along the $\langle 111\rangle$ direction. The Hamiltonian is given by
\begin{equation}\label{Model}
H = \sum\limits_{\left\langle {i,j} \right\rangle } {{K_\gamma }S_i^\gamma S_j^\gamma }  - \sum\limits_i {{\mathbf{H}} \cdot {{\mathbf{S}}_i}} .
\end{equation}
Here, $\gamma=x,y,z$ denote the three distinct nearest neighbor links $\langle {i,j}\rangle$ of the hexagonal lattice [see Fig.~\ref{Fig1}], $S^\gamma$ represents effective spin-1/2 degrees of freedom sitting on each vertex  and interacting via exchange $K_\gamma$. The ground state at $\mathbf{H}=0$ corresponds to the Kitaev limit, which exhibits two kinds of QSLs depending on the relative coupling strength. When one of the three coupling $K_\gamma$ is much larger than the others, the system is a gapped $\mathbb{Z}_2$ spin liquid with Abelian excitations, while around the isotropic point of equal couplings, the system is a gapless spin liquid\cite{Kitaev2006}. The latter can turn into a non-Abelian topological phase under time-reversal symmetry breaking perturbations \cite{Kitaev2006}, e.g., by adding a three-spin chirality term\cite{DHLee2007} or by applying an external magnetic field\cite{Kitaev2006} or by decorating the honeycomb lattice\cite{Yao2007}.

We use both exact diagonalization (ED) and density matrix renormalization group (DMRG) to study the Hamiltonian (\ref{Model}) with isotropic coupling $K_\gamma \equiv K$, as a function of an external magnetic field $\mathbf{H}$. We compare the phase diagrams with AFM ($K>0$) and FM ($K<0$) Kitaev couplings.

In the present calculation, we consider a system of  size $N=L_x\times L_y\times 2$  [see Fig.~\ref{Fig1}], where $L_x$ and  $L_y$ represent the number of unit cells along $x$ and $y$ directions, respectively. Our present DMRG calculations keep enough states to ensure the truncation error of the order or smaller than $10^{-9}$ and perform  DMRG sweeps until the measured quantities are converged.

\emph{Non-Abelian topological phase.}---We first compute the energy spectra of the model Hamiltonian Eq.~(\ref{Model}) as a function of  the magnetic field. Figure~\ref{Fig2_ED} shows  the low-energy spectra in different momentum sectors  for a system size $N= 4\times 3\times2$ on the torus, with antiferromagnetic ($K=+1$) or ferromagnetic ($K=-1$) Kitaev couplings. For the antiferromagnetic case, we find two lowest energy states in $(\pi,0)$ and $(0,0)$ momentum sectors, which are separated from the higher energy states by a finite gap for a range of magnetic field  $0\leq H_{\text{111}}\lesssim 0.2$.
In contrast, in the case with ferromagnetic Kitaev coupling, the spectra shown in Fig.~\ref{Fig2_ED} (b) indicates that the topological phase only survives in a much smaller regime at $H_{\text{111}}\lesssim 0.02$.  Meanwhile, while naively one would expect a transition directly to the partially-polarized phase (which is smoothly connected to the fully polarized $H_{\text{111}}=\infty$ limit), as is indeed seen in the FM Kitaev
model [see Fig. 2 (b), (e) and (f)], here for the AFM Kitaev model, as shown in Fig.~\ref{Fig2_ED} (a),(e) and  (f),
an intermediate gapless phase (discussed further below) is observed at $H_*^\text{AFM} < H < H_{**}^\text{AFM}$ before a transition to polarized paramagnet at $H_{**}^\text{AFM}\approx 0.36$ . In both AFM and FM cases, the critical field is also identified by sharp peaks in the second order derivative of the ground state energy or equivalently the magnetic susceptibility[see  Fig.~\ref{Fig2_ED} (f)]. Similarly to the FM case\cite{Jiang2011}, the field-driven phase transitions in the AFM case might be continuous or weakly first-order.

We now demonstrate the topological nature of the two lowest-energy states below the critical field $H_*$.  First, we note that the Kitaev model at zero field with periodic boundary conditions has topological ground state degeneracy in two dimensions. Different ground states are characterized by  two Wilson loop operators $W_y$ and $W_x$ associated with non-contractible loops along $y$ and $x$ directions respectively. As illustrated in Fig.~\ref{Fig1}, the definitions of $W_y$ and $W_x$ are given by
 \begin{align}
 {W_y} = -\left\langle {\prod\limits_{i = 1}^{2{L_y}} {\sigma_i^y} } \right\rangle;
 {W_x}  = -\left\langle {\prod\limits_{i = 1}^{2{L_x}} {\sigma_i^z} } \right\rangle.
 \end{align}
 where $\sigma^y$ and $\sigma^z$ are Pauli matrices\cite{note}, i.e.\ twice the spin--1/2 operators. The loops along $y$ direction only cover $\gamma=x,z$ links while the loops along $x$ direction only cover $\gamma=x,y$ links. It is straightforward to verify that these Wilson loop operators commute with each other and also with the Hamiltonian
in the Kitaev limit. Each operator squares to identity, hence its eigenvalue is either $+1$ or $-1$. The $\pm 1$ eigenvalue of Wilson loop operator corresponds to the $Z_2$ fluxes  or equivalently the periodic/antiperiodic boundary conditions for the emergent Majorana fermions.

The expectation values of Wilson loop operators $W_{x,y}$ are measured for these two lowest energy states in the model (\ref{Model}). As shown in Fig.~\ref{Fig2_ED} (c) and (d), these Wilson loop operators take exact quantized values in Kitaev limit and nearly quantized values for a finite range of magnetic fields below the critical value. This indicates that the emergent $\mathbb{Z}_2$ gauge theory remains a good description of perturbed Kitaev model away from static limit.  Importantly, below the critical field, the two lowest energy states have nearly the same value of $W_y \simeq -1$ but  distinct values of $W_x$, with $W_x \simeq +1$ for the state in momentum $(\pi,0)$ sector and $W_x \simeq -1$ for the state in $(0,0)$ sector, as shown in the inset of Fig.~\ref{Fig2_ED} (c). These results are fully consistent with our expectation that the degeneracy between different topological sectors of the Kitaev phase in thermodynamic limit is split by finite size effect in a quasi-one-dimensional geometry. For a three-leg system studied in this work, the splitting between $W_y = 1$ and $W_y=-1$ sectors is strong enough that the two lowest energy states both have $W_y \simeq -1$. As we shall show below, these two lowest energy states become degenerate as $L_x$ increases. Their many-body momenta $k_x=0$ and $\pi$ indicate that as a one-dimensional system the three-leg AFM Kitaev model spontaneously breaks translational symmetry breaking and doubles the unit cell in thermodynamic limit. This is analogous to the charge-density-wave states obtained by placing $\nu=1/3$ fractional quantum Hall states on a thin torus.

The expectation values of Wilson loop operators decay rapidly near the critical field and becomes negligible above the critical field. The near quantization of Wilson loop operators (and its lack of) provide another strong evidence for the topological (non-topological) nature of the phases before (after) the phase transition.

\begin{figure}[tbp]
\begin{center}
\includegraphics[width=0.45\textwidth]{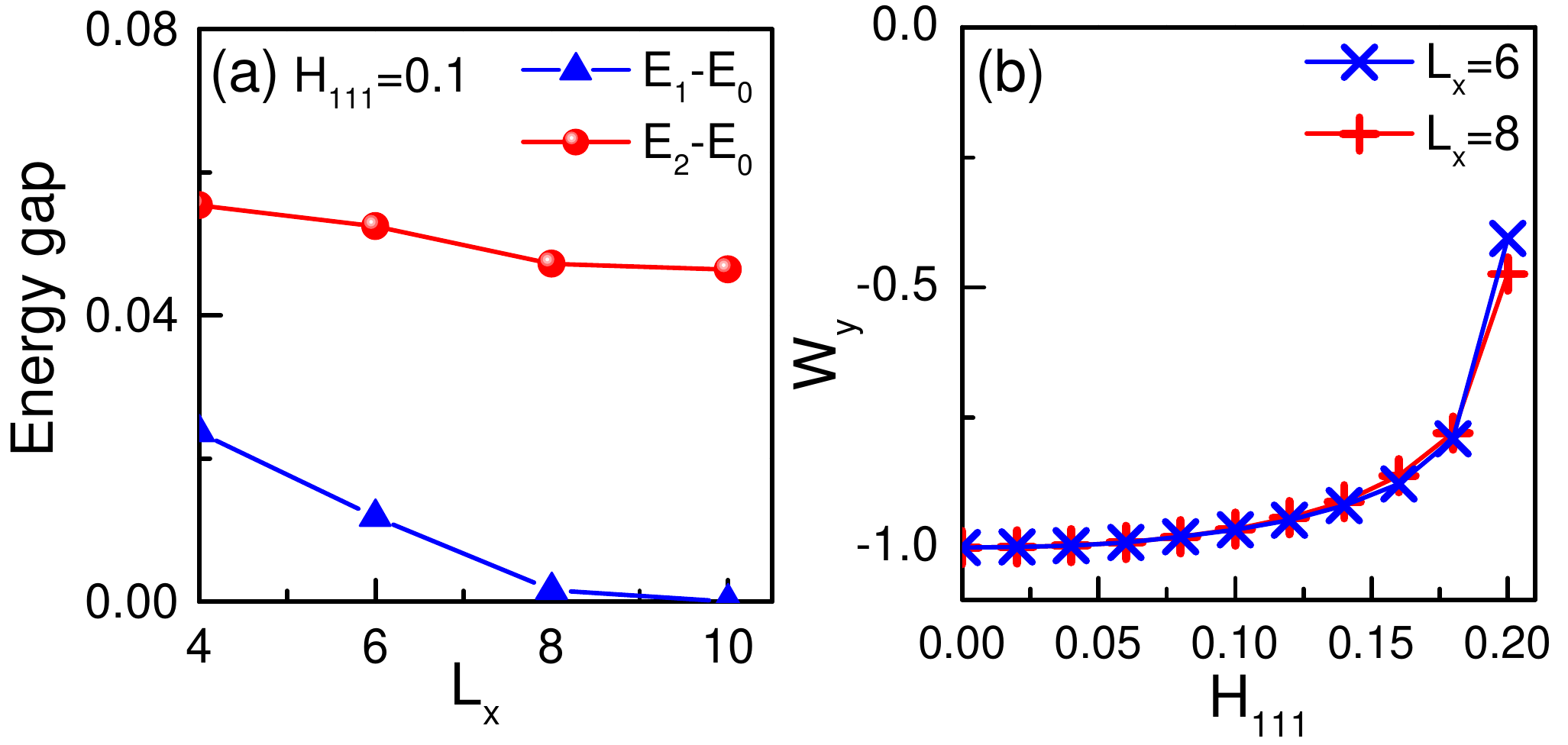}
\end{center}
\par
\renewcommand{\figurename}{Fig.}
\caption{(Color online) (a) The DMRG calculation of the lowest three energy states as function of $L_x$ at $H_{\text{111}}=0.1$ on torus. The energy difference between two lowest energy sates becomes vanishingly small with increasing system size while they are separated from higher energy sectors by a finite gap. (b)The DMRG results of the Wilson loop operator $W_y$ on torus for the antiferromagnetic (AFM) Kitaev model with $L_x=6,8$. }
\label{Fig3}
\end{figure}

For the AFM Kitaev model, we further use DMRG to calculate the ground state degeneracy at $H_{\text{111}}\lesssim 0.2$ for large $L_x$ to confirm its topological nature.  In Fig.~\ref{Fig2_ED} (a), we find a small energy split between  two lowest energy states. To confirm these two states are exactly degenerate states in the thermodynamic limit, we perform DMRG calculation on torus by targeting three lowest  energy states with increasing $L_x$. As shown in Fig.~\ref{Fig3} (a), we find that the energy difference between two lowest states $E_1-E_0$ becomes vanishingly small when the system length $L_x \gtrsim 8$ , while the lowest two states are separated from higher energy states by a finite gap indicated by $E_2-E_0$ [see Fig.~\ref{Fig3} (a)].  Based on these calculation, the two-fold ground state degeneracy of such topological phase is identified. Meanwhile, we also checked the Wilson loop operator $W_y$ for different $L_y=3$ system size by DMRG, as shown in  Fig.~\ref{Fig3} (b), we find the topological phase is very robust and independent of system length.

\begin{figure}[tbp]
\begin{center}
\includegraphics[width=0.45\textwidth]{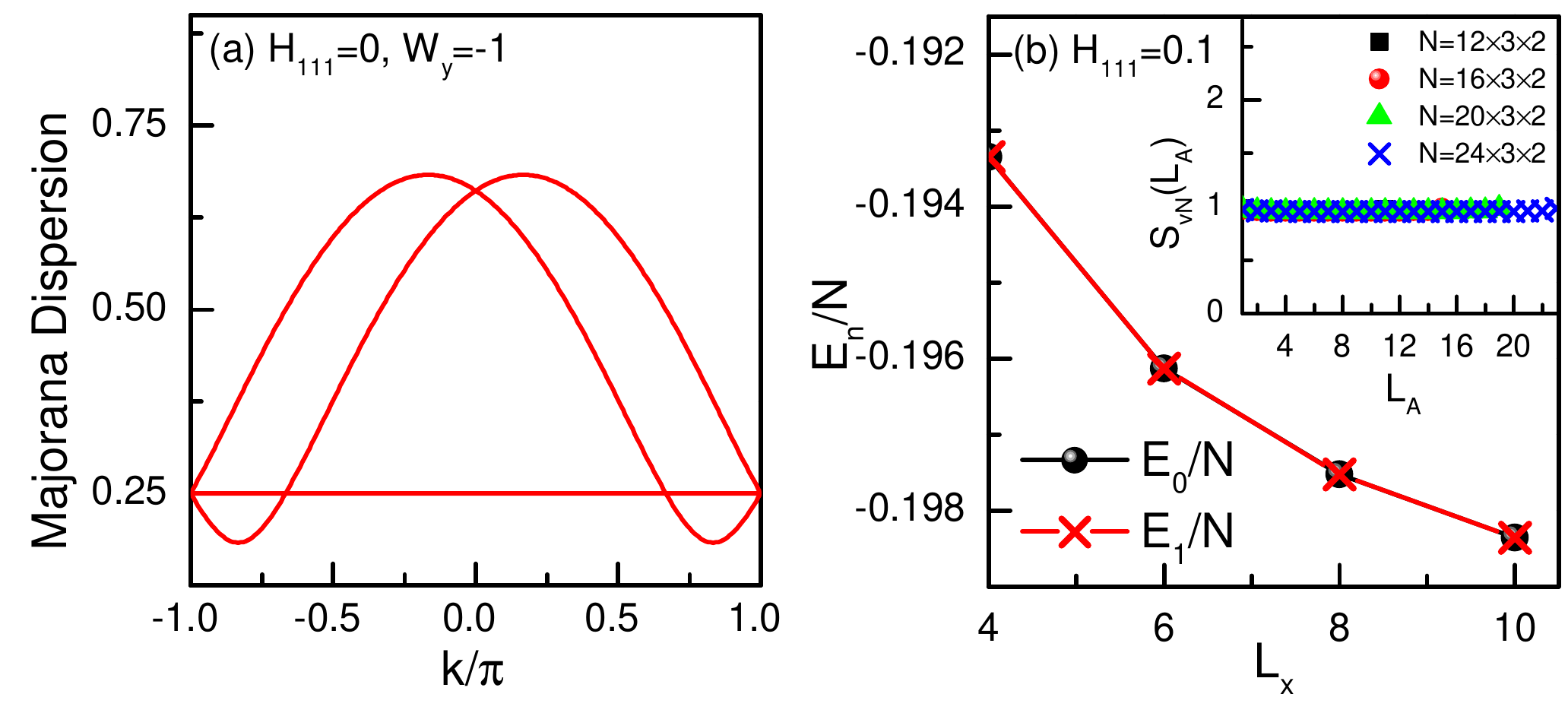}
\end{center}
\par
\renewcommand{\figurename}{Fig.}
\caption{(Color online) (a)Majorana dispersions on infinite ladders ($L_y=3$) for $W_y=-1$ sector at $H_{\text{111}}=0$,  which show finite gap and quantized value of Wilson loop operators.  These are consistent with the numerics for finite fields (see main text). (b) The energy density of two degenerate states on cylinders at $H_{\text{111}}=0.1$, the inset shows the Von Neumann entanglement entropy  for  long cylinders, where the flat feature indicates the existence of the finite gap. }
\label{Fig:Band}
\end{figure}

The gapped feature of the topological phase can also be confirmed by the Von Neumann entanglement entropy $S_\text{VN}$ defined by ${S_\text{VN}} =  - Tr\left( {{\rho _\text{A}}\ln {\rho _\text{A}}} \right)$, where ${\rho _\text{A}}$ is the reduced density matrix of part $A$ for the bipartition of the system into $A$ and $B$, ${\rho _\text{A}}$ is got by tracing out the degrees of freedom of $B$ part. Here, we consider the cut parallel to $y$ direction and measure  the value of $S_\text{VN}$ for each cut at $L_A$. For the gapped state, the Von Neumann entropy should be independent on the positions of each cut and display flat behavior. As shown in Fig.~\ref{Fig:Band} (b), we calculate a long cylinder by DMRG and find the flat  $S_\text{VN}$ as a function of $L_A$, implying the existence of the well defined gap in the topological phase. All of these confirm the stability of the topological phase.

In the absence of magnetic field, the Kitaev model is exactly solvable in terms of static fluxes and Majorana fermions\cite{Kitaev2006}. We also analyze the exact solution  on finite systems as well as infinite ladders. In each topological sector defined by a particular set of values of the Wilson loop operators, the ground state energy is simply given by the energy of the filled fermi sea, i.e.\ the sum of all negative Majorana eigenvalues. Figure~\ref{Fig:Band} (a) shows the Majorana dispersion  for infinite cylinders in $W_y=-1$ sector with fixed width $L_y=3$.  We also compared the exact solution on finite-sized systems with DMRG and ED results, which are consistent with each other.

Interestingly, the three-leg system with $W_y=-1$ as we identified here is a one-dimensional topological superconductor of Majorana fermions in the thermodynamic limit $L_x \rightarrow \infty$. This is shown by computing $\text{sgn}[\text{Pf}[H[0]]\text{Pf}[H[\pi]]]$, i.e.\ the sign of the product of Pfaffians of the quadratic Majorana Hamiltonian matrices at 1D momenta $k=0$ and $k=\pi$.  We find a negative value for this topological index, correspond to a 1D topological superconductor\cite{Kitaev2000}. Therefore, we expect the presence of boundary Majorana zero modes for open boundary conditions in the $L_x$ direction.
These boundary zero modes can be regarded as a descendent of non-Abelian anyons in the Kitaev phase in two dimensions, and their presence should be robust against perturbations such as the magnetic field.
Indeed, we find the lowest two states are exactly degenerate on cylinders by DMRG, as shown in the Fig.~\ref{Fig:Band} (b) for $H_{\text{111}}=0.1$, and the  two-fold degeneracy in the entanglement spectrum on cylinders, confirming the existence of Majorana zero modes on the boundary.

\emph{Discussion and Summary.}---In this letter, we report a robust non-Abelian phase in the antiferromagnetic Kitaev model under magnetic field. Based on extensive DMRG and ED simulations, we identify its topological features by the energy spectra, entanglement, topological degeneracy, and  Wilson loop operators. We find that the topological phase in the antiferrmagnetic Kitaev model is much more stable to increasing magnetic field than the one in ferromagnetic Kitaev model.
This can be partially understood from the low field magnitude of magnetic susceptibility (Fig.~\ref{Fig2_ED} (a,b) insets), which in turn have a simple interpretation. While at zero field the AFM and FM Kitaev models are exactly equivalent by a majorana sign transformation on one honeycomb sublattice\cite{Kitaev2006},
since their spin correlations are identical except opposite in sign, the FM Kitaev model is nearly a ferromagnet, while the AFM model has similar strong response to a staggered magnetic field but a weak response to a uniform field.
This difference between the AFM and FM Kitaev coupling can also be seen approaching from the infinite field limit\cite{Janssen2016} based on semiclassical spin wave analysis.
Our findings suggest that, in materials with dominant antiferromagnetic Kitaev interactions, a spin liquid phase if present may be observable under application of fairly substantial magnetic fields, in contrast to previous expectations.

Moreover, when the gapped chiral topological order is destroyed by the large field in the AFM case, before entering the polarized phase an intermediate gapless phase is found. Within the intermediate gapless phase of the AFM model, the DMRG algorithm converges to a state exhibiting modulations in spin density  around the partially-polarized mean (about $10\%\sim 20\%$ of full amplitude), which  appear to be pinned by the open boundaries (see Supplementary Material \cite{supplementary} for details). Together with the gapless spectrum [Fig.~\ref{Fig2_ED} (a)] and the large entanglement,  these observations serve as evidence that this gapless  phase involves long range correlations or entanglement, and thus it cannot be captured reliably in the 2D limit. A possible connection to experiments remains an open question.

\begin{acknowledgments}
\emph{Acknowledgments}---We would like to thank James Analytis, Radu Coldea, and Yang Qi for insightful discussions. Z.Z.  and L.F. are supported by the David and Lucile Packard foundation. I.K. is supported by the MIT Pappalardo Fellowship.  D.N. Sheng is supported by the U.S. Department of Energy, Office of Basic Energy Sciences under grants No. DE-FG02-06ER46305.  Z.Z. used the Extreme Science and Engineering Discovery Environment (XSEDE) to perform part of simulation, which is supported by National Science Foundation grant number ACI-1548562.
\end{acknowledgments}

\onecolumngrid
\newpage
\renewcommand{\theequation}{S\arabic{equation}}
\setcounter{equation}{0}
\renewcommand{\thefigure}{S\arabic{figure}}
\setcounter{figure}{0}
\renewcommand{\bibnumfmt}[1]{[S#1]}
\begin{center}
 {\bf {Robust non-Abelian spin liquid and possible intermediate phase in antiferromagnetic Kitaev model with magnetic field: Supplementary Material}}
\end{center}

In the main text,  we  mainly focus on the robustness  of topological phase against external magnetic field H along the 111-direction.  Based on systematically numerical simulations by density matrix renormalization group (DMRG) and exact diagonalization (ED), we identify the remarkable stability  of the topological phase in the antiferromagnetic (AFM)  Kitaev model by the ground state degeneracy on the torus and the Wilson loop operators. In this supplementary material, we will provide additional numerical evidence to address the interesting physics in this model, including the field driven transitions and the nature of the intermediate gapless phases.

\section{ The field-driven phase transitions }
\begin{figure}[htbp]
\begin{center}
\includegraphics[width=0.9\textwidth]{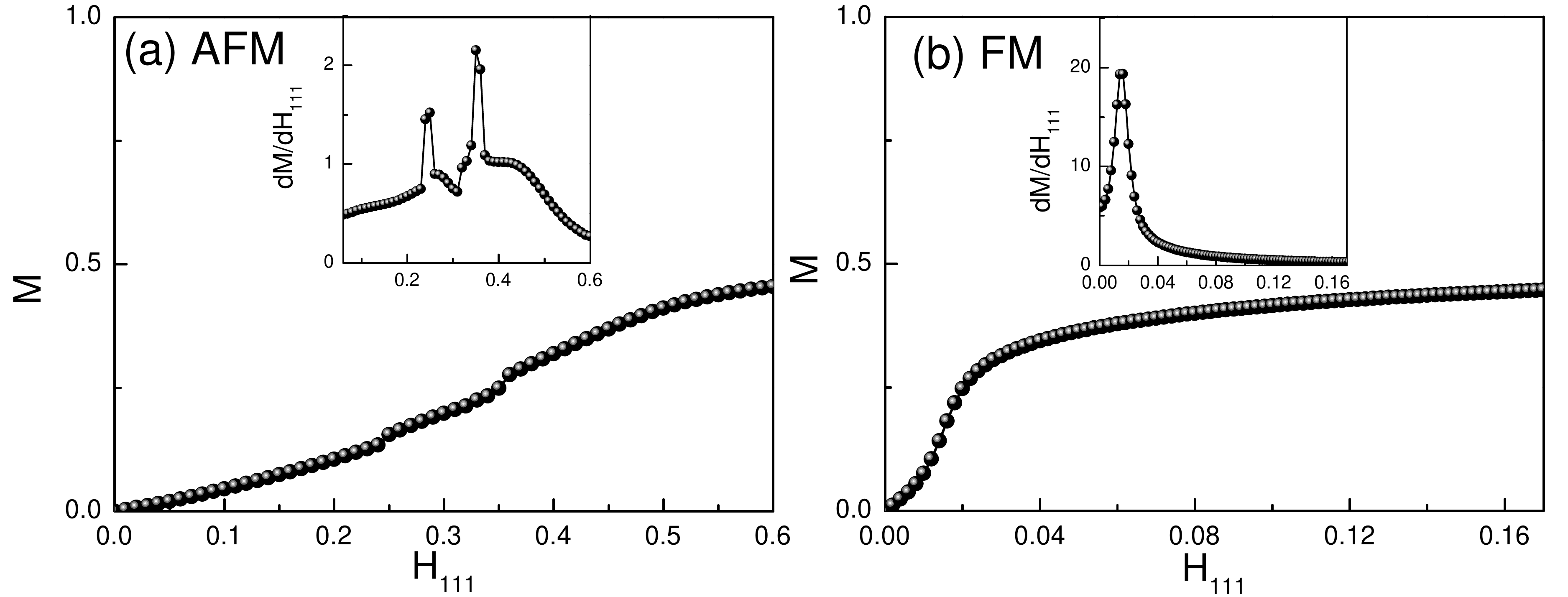}
\end{center}
\par
\renewcommand{\figurename}{Fig.}
\caption{(Color online) The magnetization curve  and its first order derivative curve (insets) for the antiferromagnetic  (a) and ferromagnetic (b) Kitaev model.  Here,  the system size $N= 3\times 3 \times2$ and we apply fully periodical boundary conditions along two directions.}
\label{FigS1}
\end{figure}

In the main text we compared the stability of the topological phase in the AFM Kitaev model and FM Kitaev model. While the FM Kitaev spin liquid is destroyed by a weak magnetic field, the AFM spin liquid remains robust up to a magnetic field that is an order of magnitude larger. {Here, we noted that, in the Kitaev limit, the energy spectrum in Fig.~2 in the main text indicates the Majorana gap is larger than the pi-flux gap on finite sized system.} In addition, an intermediate gapless phase is observed in the AFM Kitaev model for larger fields, before a second transition to the high-field partially-polarized paramagnet. The transition among different phases can be determined by the peaks in the second order derivative of ground state energy or the magnetic susceptibility [Fig.~2 (f) in the main text]. Moreover, the phase boundaries can also be identified by the magnetization curve [Fig.~2 (e) and (f) in the main text], with the magnetization defined by
\begin{equation}
M \equiv \frac{1}{N}\sum\limits_i {\left\langle {S_i^x + S_i^y + S_i^z} \right\rangle/\sqrt{3} }.
\end{equation}
The magnetization curve is experimentally relevant and its first order derivative should be the magnetic susceptibility. Figure ~\ref{FigS1} shows the the magnetization curve of the same model with both the AFM Kitaev coupling [see Fig~\ref{FigS1} (a)] and the FM Kitaev coupling [see Fig~\ref{FigS1} (b)], we find there is no significant discontinuity in the curve, and the first order derivative curves  [see the insets in Fig~\ref{FigS1} (a) and (b)] exhibit significant peaks near the critical points. These numerical data suggest that the field-driven phase transitions might be continuous or weakly first order.

\begin{figure}[htbp]
\begin{center}
\includegraphics[width=0.9\textwidth]{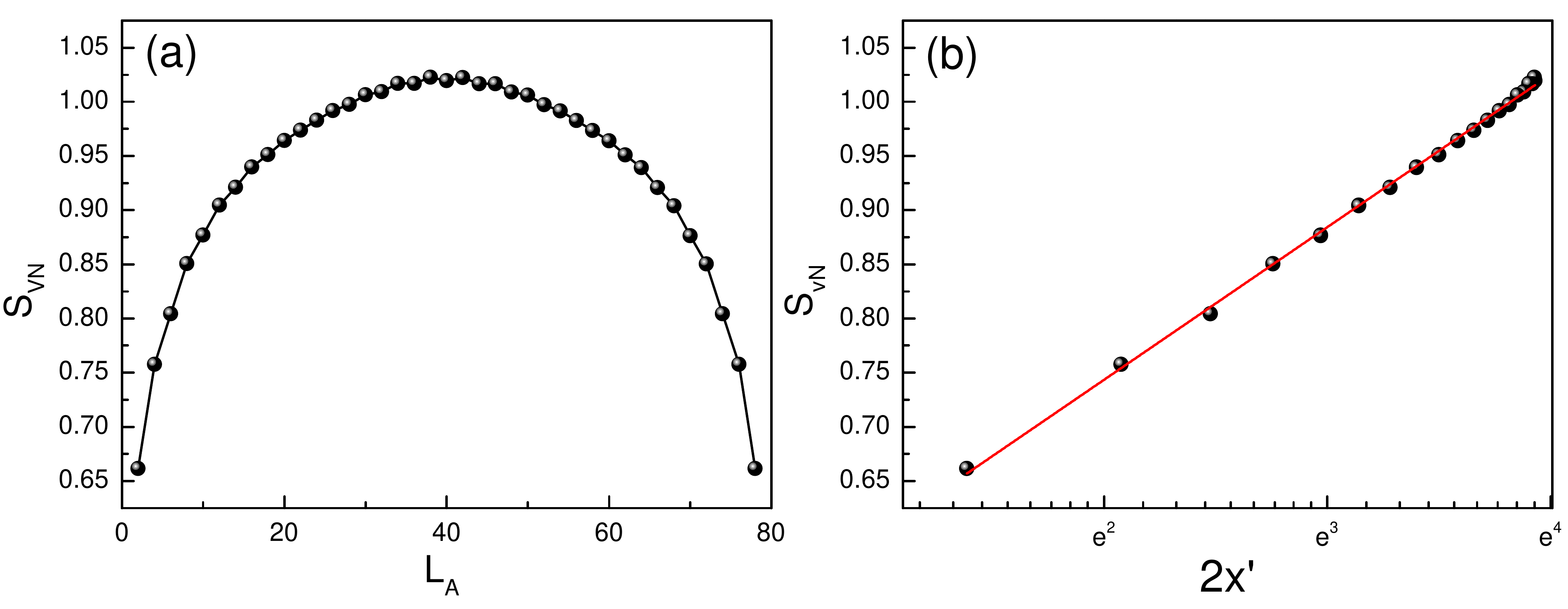}
\end{center}
\par
\renewcommand{\figurename}{Fig.}
\caption{(Color online) The von Neumann entanglement entropy $S_\text{VN}$ of the intermediate phase on cylinders. (a) $S_\text{VN}$ as a function of   each cut at $L_A$, which is parallel to $y$ direction; here $L_A$ is an even number considering the even-odd effect when applying open boundary condition along $x$ direction;  (b) $S_\text{VN}$ as a function of  the conformal distance  $x'$=$(L/{\pi})sin({\pi L_A}/L)$. Here, the system size $N= 80\times 2 \times2$. The fitting of the curve in (b) shows the central charge got from the calculation is close to $c=1$. }
\label{Fig:Entropy}
\end{figure}

\begin{figure}[htbp]
\begin{center}
\includegraphics[width=0.95\textwidth]{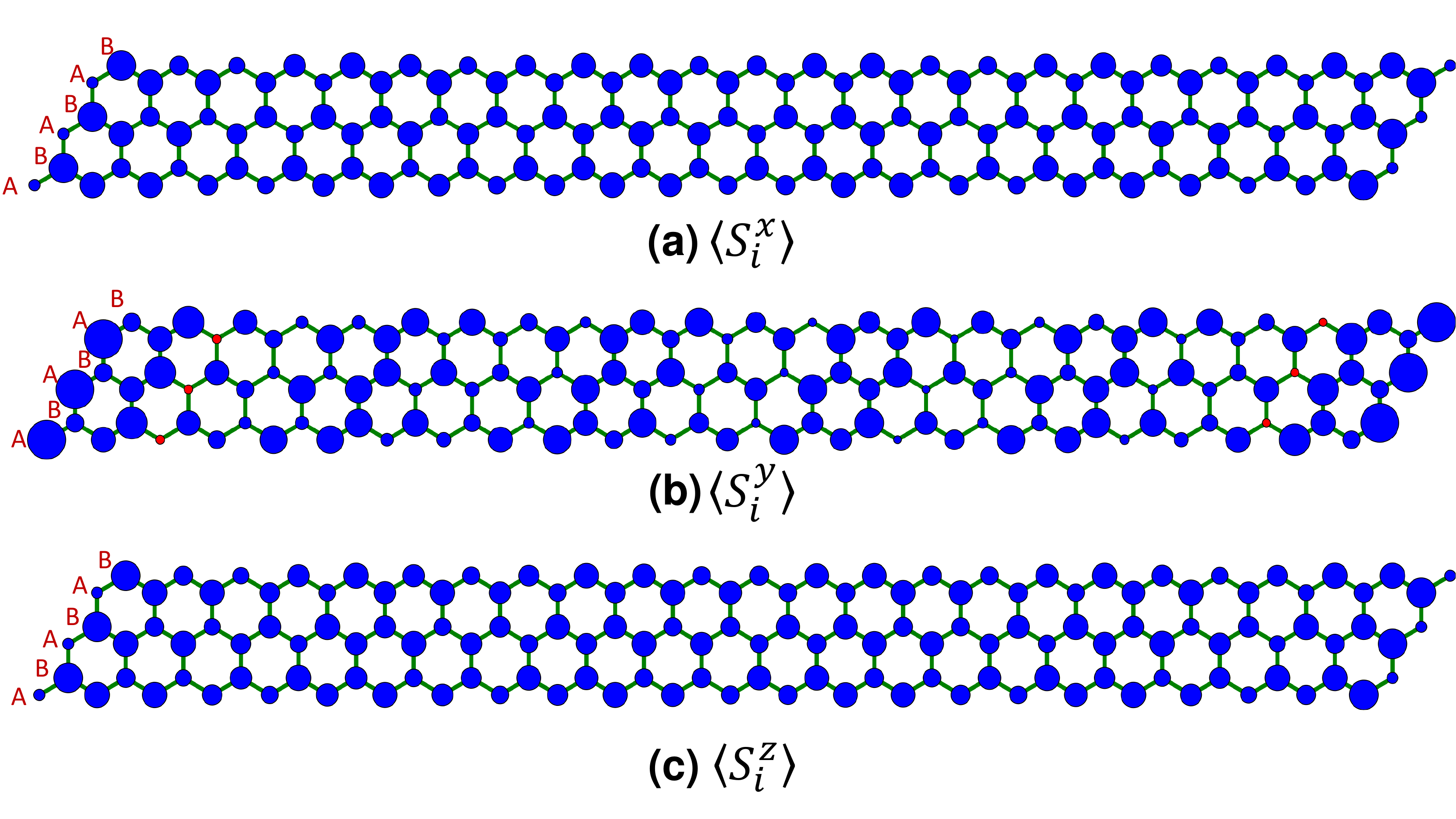}
\end{center}
\par
\renewcommand{\figurename}{Fig.}
\caption{(Color online) The spin density distribution pattern for antiferromagnetic Kitaev model at $H_{\texttt{111}}=0.3$ (a) $\left\langle S_i^x  \right\rangle$; (b) $\left\langle S_i^y  \right\rangle$; (c) $\left\langle S_i^z  \right\rangle$.  The area is proportional to the absolute value of the spin density, while the blue (red) color represents the value is positive (negative). Here,  the system size $N= 24\times 3 \times2$ and we apply cylinder boundary conditions.}
\label{Fig:SpinDensity}
\end{figure}

\begin{figure}[htbp]
\begin{center}
\includegraphics[width=0.9\textwidth]{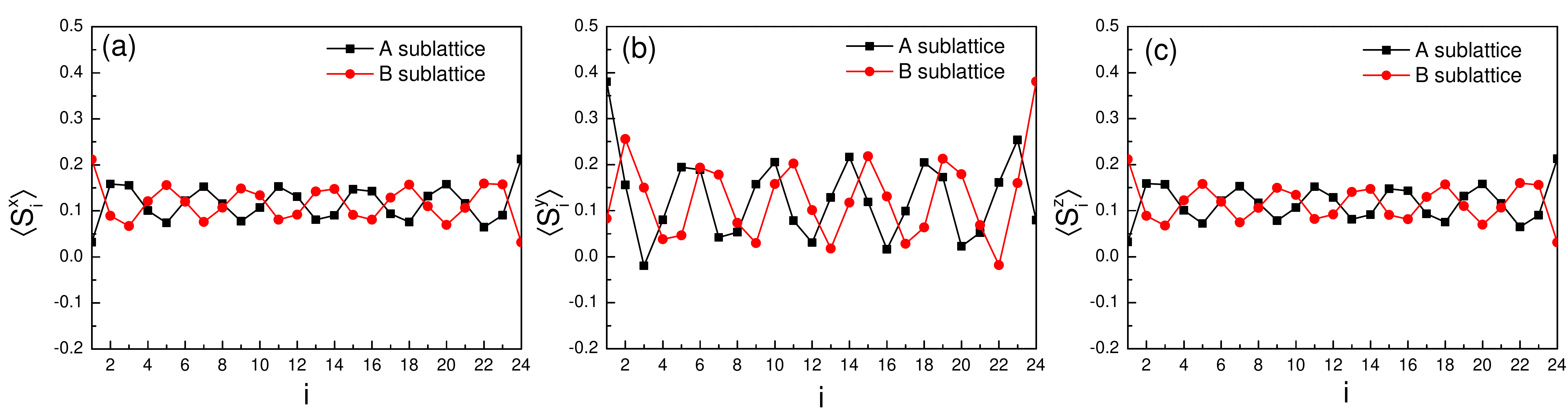}
\end{center}
\par
\renewcommand{\figurename}{Fig.}
\caption{(Color online) The spin density distribution for antiferromagnetic Kitaev model along the $x$ direction at $H_{\texttt{111}}=0.3$ (a) $\left\langle S_i^x  \right\rangle$; (b) $\left\langle S_i^y  \right\rangle$; (c) $\left\langle S_i^z  \right\rangle$.  Here,  the system size $N= 24\times 3 \times2$ and we apply cylinder boundary conditions.}
\label{Fig:SpinDensity2}
\end{figure}

\section{ The intermediate phase in the AFM  Kitaev model}

For the Kitaev model in an external magnetic field, one would expect a direct transition from the topological phase to a  partially-polarized phase (which is smoothly connected to the fully polarized $H_{\text{111}}=\infty$ limit), as is indeed seen in the FM Kitaev model [see Fig. 2 (b) in the main text]. However, an intermediate phase is observed  in the AFM Kitaev model, before a second transition to the high-field partially-polarized paramagnet, when we study the energy spectra  as a function of  the magnetic field. The energy spectra in the Fig. 2 (a) in the main text indicates the intermediate phase is gapless. Below we present additional numerical evidence to gain some hints on its nature.

Firstly, we  study the  Von Neumann entanglement entropy $S_\text{VN}$. Considering a bipartition of the system into parts $A$ and $B$, and the Hilbert space can be written as a direct product $H$ = ${H_A} \otimes {H_B}$. Then the reduced density matrix of $A$  (i.e., ${\rho _\text{A}}$) is obtained by tracing out the degrees of freedom of $B$ part, i.e.,  ${\rho _A} = {\text{Tr}_\text{B}}\rho $, where$\rho  = \left| {{\psi _0}} \right\rangle \left\langle {{\psi _0}} \right|$ is the density matrix of ground state $\left| {{\psi _0}} \right\rangle $. The Von Neumann entropy  is defined as
\begin{equation}
 {S_\text{VN}} =  - Tr\left( {{\rho _\text{A}}\ln {\rho _\text{A}}} \right).
 \end{equation}
Here, we consider the cut parallel to $y$ direction and measure  the value of $S_\text{VN}$ for each cut at $L_A$.  On finite sized systems, one can use the conformal mappings $L_A \rightarrow x'=(L/{\pi})sin({\pi L_A}/L)$ for periodical boundary condition (PBC), and $L_A \rightarrow 2x'$ for open boundary condition (OBC).  Within the CFT\cite{CFT1,CFT2,CFT3},
\begin{equation}
 {S_\text{VN}}= \frac{c}{3}\ln \left( {x'} \right) + {S_1}
 \end{equation}
 for PBC, where  $c$ is the central charge , and
 \begin{equation}\label{CentralOBC}
 {S_\text{VN}}= \frac{c}{6}\ln \left( {2x'} \right) + \ln \left( g \right) + {S_1}/2
   \end{equation}for OBC. Here, $S_1$ is a model dependent constant, and $g$ is a universal boundary term\cite{CFT3}.

For the gapped state, the von Neumann entanglement entropy $S_\text{VN}$ should be independent on the positions of each cut and displays flat behavior, such as $S_\text{VN}$ in the topological phase shown in the Fig. 4(b) in the main text. Figure~\ref{Fig:Entropy}  shows  $S_\text{VN}$ of the intermediate phase on the cylinder. The profile in Fig.~\ref{Fig:Entropy} (a) also indicates the intermediate phase is gapless with finite central charge, which is consistent with the gapless feature in the energy spectra got by ED.  After the conformal mappings, we fit the central charge based on the Eq.~\ref{CentralOBC}, the fitting indicates the central charge $c\approx1$ [see Fig.~\ref{Fig:Entropy} (b) ].

 From the energy spectrum and the entanglement entropy, we can find the gapless nature of the intermediate phase. Here, it should also be noted that  the numerical calculation of the current model is already a great challenging endeavor, not only because one has to deal with the entire Hilbert space due to the lack of SU(2) and U(1) (any of spin components is not conserved) symmetry, but also because one has to use complex data types in the DMRG code due to the 111-orientation of the magnetic field. The gapless nature further increase the computational cost of DMRG due to the high entanglement. This is because the computational cost of DMRG is determined by the entanglement, to simulate a system with a lot of entanglement, the bond dimension (and thus the memory and time of the computation) grows exponentially with the entropy, which further increases the numerical complexity.

 We measured the  spin density distribution along the cylinders, as shown in the Fig.~\ref{Fig:SpinDensity} and Fig.~\ref{Fig:SpinDensity2}. The three components of spin densities are uniform for the same sublattices  along $y$ direction  [see Fig.~\ref{Fig:SpinDensity}] due to the periodical boundary condition, while it displays spatial modulations along the cylinder with   $10\%\sim 20\%$ of full amplitude and involves a range of wave vectors pinned by the open boundaries, as shown in  Fig.~\ref{Fig:SpinDensity2}.  Considering the gapless nature of the intermediate phase, these observations indicate that this gapless  phase involves long range correlations or entanglement, and thus it cannot be captured reliably in 2D limit by DMRG, which remains an open question.


\begin{thebibliography}{99}
\bibitem{note}The necessity of the minus sign in these formulas can be seen by considering the $Z_2$ gauge fields along the path of the Wilson loop traversing the periodic boundary conditions.
\end{thebibliography}
\end{document}